\UseRawInputEncoding

\documentclass[aps,showpacs,twocolumn,pra,superscriptaddress]{revtex4}

\usepackage[tbtags]{amsmath}
\usepackage{amssymb}
\usepackage{amsfonts}
\usepackage{bmpsize}
\usepackage{mathtools}
\usepackage[colorlinks=true,citecolor=blue,urlcolor=blue]{hyperref}
\hypersetup{colorlinks,linkcolor=blue,filecolor=blue,urlcolor=blue,citecolor=blue,}
\usepackage{multirow}

\begin{document}
\title{Non-locality sharing for a three-qubit system via multilateral sequential measurements}

\author{Changliang Ren}\thanks{Corresponding author: renchangliang@hunnu.edu.cn}
\affiliation{Key Laboratory of Low-Dimensional Quantum Structures and Quantum Control of Ministry of Education, Key Laboratory for Matter Microstructure and Function of Hunan Province, Department of Physics and Synergetic Innovation Center for Quantum Effects and Applications, Hunan Normal University, Changsha 410081, China}
\author{Xiaowei Liu}
\affiliation{Key Laboratory of Low-Dimensional Quantum Structures and Quantum Control of Ministry of Education, Key Laboratory for Matter Microstructure and Function of Hunan Province, Department of Physics and Synergetic Innovation Center for Quantum Effects and Applications, Hunan Normal University, Changsha 410081, China}
\author{Wenlin Hou}
\affiliation{Key Laboratory of Low-Dimensional Quantum Structures and Quantum Control of Ministry of Education, Key Laboratory for Matter Microstructure and Function of Hunan Province, Department of Physics and Synergetic Innovation Center for Quantum Effects and Applications, Hunan Normal University, Changsha 410081, China}
\author{Tianfeng Feng}
\affiliation{State Key Laboratory of Optoelectronic Materials and Technologies and School of Physics, Sun Yat-sen University, Guangzhou, People's Republic of China}
\author{Xiaoqi Zhou}\thanks{Corresponding author: zhouxq8@mail.sysu.edu.cn}
\affiliation{State Key Laboratory of Optoelectronic Materials and Technologies and School of Physics, Sun Yat-sen University, Guangzhou, People's Republic of China}
\begin{abstract}
Non-locality sharing for a three-qubit system via multilateral sequential measurements was deeply discussed. Different from 2-qubit cases, it is shown that non-locality sharing between $\mathrm{Alice_{1}-Bob_{1}-Charlie_{1}}$ and $\mathrm{Alice_{2}-Bob_{2}-Charlie_{2}}$ in a 3-qubit system can be observed, where two Mermin-Ardehali-Belinskii-Klyshko (MABK) inequalities can be violated simultaneously. What¡¯s more, a complete non-locality sharing with all of 8 MABK inequalities simultaneous violations can be also observed. Compared with 2-qubit cases, the nonlocal sharing in a 3-qubit system shows more novel characteristics. Finally a general interpretation of nonlocal sharing according to joint conditional probability was discussed.
\end{abstract}


\maketitle

\section{Introduction}

Local realism indicates the nature of the world that measurement outcomes are pre-deterministic, and the measurement on one party of a multipartite system does not affect the other parties. However, quantum mechanics predicts that there are stronger correlations than the correlations of local hidden variables because of inherent non-locality of quantum theory \cite{Einstein}. The so-called Bell inequality is exploited to distinguish the differences between classical correlation and quantum correlation \cite{Bell}. Subsequently, Bell-type inequalities have been studied extensively from various perspectives \cite{Clauser,Zukowski,Mermin,Belinskii,Ardehali,Collins,Brukner, Lee} and experimentally verified in many different quantum systems \cite{Aspect,Weihs,Rowe,Hofmann,Giustina,Christensen,Hensen,Giustina1,Shalm}. These kinds of research, are not only important to deeply understand quantum theory, but also play an crucial role in quantum information protocols, such as quantum key distribution \cite{Acin}, randomness generation\cite{Colbeck,Pironio,Pirandola, Liu, Bierhorst}, and entanglement certification\cite{Bowles}. For a background on Bell inequalities, readers could refer to \cite{Brunner}, and references therein.

Inspired by Bell's work, Clauser, Horne, Shimony, and Holt (CHSH) derived a modified inequality \cite{Clauser}, which provides a faithful way for experimentally testing the non-locality property in 2-qubit composite systems. However, most discussions of non-locality based on CHSH inequality focus on one pair of entangled qubits distributed to only two separated observers. Recently, a surprising result that non-locality can actually be shared among more than two observers using weak measurements, has been reported by Silva \emph{et al. }\cite{Silva}. In Silva's scenario, a pair of maximally-entangled qubits is distributed to three observers Alice, $\mathrm{Bob_1}$ and $\mathrm{Bob_2}$, in which Alice accesses one qubit and the two Bobs access the other qubit. Alice performs a strong measurement on her own qubit, while $\mathrm{Bob_1}$ performs a weak measurement on his qubit and passes it to $\mathrm{Bob_2}$. Finally $\mathrm{Bob_2}$ carries out a strong measurement. The measurement results reveal that it is possible to observe a simultaneous violation of CHSH inequality in Alice-$\mathrm{Bob_1}$ and Alice-$\mathrm{Bob_2}$. To date, a series of fruitful related theoretical researches \cite{DAS,Mal,Sasmal,Bera,Datta,Shenoy,Kumari,Ren,Saha,Mohan,Roy,Srivastava,Kanjilal,Yao,Zhu,Cheng} have been proposed by tracking this path and several experimental demonstrations have also been performed \cite{Hu,Schiavon,Feng}. Especially, in Ref.\cite{Ren,Feng}, it shows a observation of non-locality sharing in a wide area even if $\mathrm{Bob_1}$'s measurement is close to a strong measurement, which is impossible in the original protocol \cite{Silva}. Nevertheless, almost all discussions are limited to one-sided sequential case, i.e., one entangled pair is distributed to one Alice and multiple Bobs. Recently, Zhu \emph{et.al} explored the non-locality sharing in two-sided sequential measurements case in which one entangled pair is distributed to multiple Alices and Bobs \cite{Zhu}. But non-locality sharing between $\mathrm{Alice_1-Bob_1}$ and $\mathrm{Alice_2}-\mathrm{Bob_2}$ is impossible in such scenario \cite{Zhu}. In this letter, we explored the non-locality sharing for a three-qubit system via multilateral sequential measurements. Different from 2-qubit cases, it is shown that non-locality sharing between $\mathrm{Alice_{1}-Bob_{1}-Charlie_{1}}$ and $\mathrm{Alice_{2}-Bob_{2}-Charlie_{2}}$ in 3-qubit system can be observed, where two MABK inequalities can be violated simultaneously. Not only that,a complete non-locality sharing with all of 8 MABK inequalities simultaneous violations can be also observed. These results not only shed new light on the interplay between non-locality and quantum measurements, especially the emergence of non-locality sharing via weak measurements, but also can be applied in unbounded randomness certification \cite{Curchod}, quantum coherence \cite{Datta} and quantum steering \cite{Shenoy}.

\begin{figure}[htbp]\label{Scenario}
      \centering
      \includegraphics[width=0.4\textwidth]{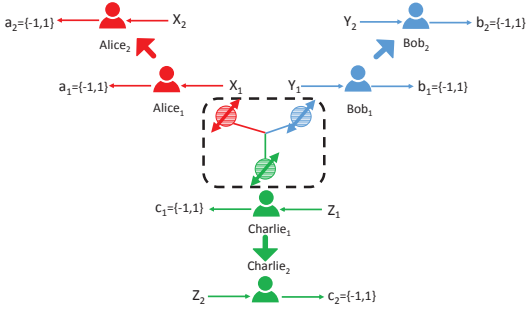}
      \caption{\small{Scenario of three-sided sequential case: a 3-qubit entangled state is distributed to three sides, and each side has two observers, in which the first observers perform weak measurements and the second perform strong measurements. Two Alices occupy one-third of the state and Bobs, Charlies have access to the different one-third after Alices' measurements. Here $\hat{X_{i}}, \hat{Y_{j}}, \hat{Z_{k}}$ represents measurement directions where $\{i,j,k\}\in\{1,2\}$ and $a_i ,b_j, c_k$  represent the results. Each observer has two directions and each measure direction has two outcomes which are $\{1,-1\}$.}} \label{2bob}
    \end{figure}

\section{Nonlocal scenario in a 3-qubit system via multilateral sequential Measurements}

 The scenario illustrated in Fig.1 is considered, where a 3-qubit entangled state is distributed to three remote sides, and each side has two observers, which can be named as $\{\mathrm{Alice_{1}, Alice_{2}}\}$,$ \{\mathrm{Bob_{1}, Bob_{2}}\}$ and $\{\mathrm{Charlie_{1},  Charlie_{2}}\}$ respectively. Those observers in the same side will measure their shared qubit sequentially. The communication is forbidden between the observers, and the measurement choice of these observers are independent. Each observer randomly chooses one of two observables to measure, which can be defined as $\hat{X_{i}}=A_{i,l}$ for $\mathrm{Alice_{i}}$, $\hat{Y_{j}}=B_{j,m}$ for $\mathrm{Bob_{j}}$ and $\hat{Z_{k}}=C_{k,n}$ for $\mathrm{Charlie_{k}}$ respectively, where $A_{i,l}$ is the $l$-th measurement chosen by the observer $\mathrm{Alice_{i}}$, similar definition for $B_{j,m}$ and $C_{k,n}$, $\{i,j,k\}\in\{1,2\}$ and  $\{l,m,n\}\in\{1,2\}$. The binary outcomes of the dichotomic measurement for these observers are given by $a_{i}$, $b_{j}$, $c_{k}$ with $\{a_{i},b_{j},c_{k}\}\in \{-1,1\}$. Such a scenario is characterized by the joint conditional probabilities of the outcomes $P(a_{1},a_{2},b_{1},b_{2},c_{1},c_{2}\mid \hat{X}_{1},\hat{X}_{2},\hat{Y}_{1},\hat{Y}_{2},\hat{Z}_{1},\hat{Z}_{2} )$.

 In the scenario, the first observer of each side performs the optimal weak measurements, while the second observer of each side will carry out strong measurements. We assumed that they share
a 3-qubit state, and the density matrix is defined as $\rho$. In the whole measurement process, we can always obtain the quantum state after measurement according to the selection of measurement and it's outcome. Without loss of generality, $\mathrm{Alice_1}$ firstly performs a weak measurement $\hat{X}_1$ on her received qubit with the quality factor $F_1$ and precision factor $G_1$ of the measurement, where $ \hat{X}_i=A_{i,l}= \vec{\xi}_{i,l} \cdot \vec{\sigma}$, $\vec{\sigma}$ is a vector consists of three Pauli matrices $\vec{\sigma}=(\sigma_x,\sigma_y,\sigma_z)$
and $ \vec{\xi}_{i,l} $ is a direction vector on the Bloch sphere, $\vec{\xi}_{i,l}=(\sin\theta_{i,l}\cos\phi_{i,l},\sin\theta_{i,l}\sin\phi_{i,l},\cos\theta_{i,l})$. As is introduced in \cite{Silva}, there exists a trade-off between measurement disturbance and information gain, where the optimal weak measurement requires that $F^2_1+G^2_1 = 1$, which means that more information can be extracted with same disturbance. We assume that the weak measurement process in this scenario is the optimal pointer case. When the measurement outcome is $a_i$, according to the discussion in \cite{Silva}, the state changes to
\begin{align}
\rho_{\hat{X}_1}^{a_1}=&\frac{F_1}{2}\rho+\frac{1+a_1G_1-F_1}{2}[U_{\hat{X}_1}^{-1}\rho(U_{\hat{X}_1}^{-1})^ \dagger]\nonumber \\ &+\frac{1-a_1G_1-F_1}{2}[U_{\hat{X}_1}^{+1} \rho(U_{\hat{X}_1}^{+1})^ \dagger]
\end{align}
 where $U_{\hat{X}_i}^{a_i}=\Pi_{\hat{X}_i}^{a_i}\otimes I\otimes I$ and  $\Pi_{\hat{X}_i}^{a_i}=\frac{I+a_i\hat{X}_i}{2}$. Subsequently, if $\mathrm{Alice_2}$ performs a strong measurement $\hat{X}_2$ with the outcome $a_2$, the 3-qubit state will change to
\begin{align}
\rho_{\hat{X}_2}^{a_2}=U_{\hat{X}_2}^{a_2}\rho_{\hat{X}_1}^{a_1} {U_{\hat{X}_2}^{a_2}}^ \dagger.
\end{align}

Then, suppose that the measurements of Bobs are later than that of $\mathrm{Alice_2}$, similarly the measurements of $\mathrm{ Charlies}$ are later than that of $\mathrm{Bob_2}$. $\mathrm{Bob_1}$ performs a weak measurement $\hat{Y}_j$ on his received qubit with the quality factor $F_2$ and precision factor $G_2$ of the measurement, where the optimal weak measurement requires that $F^2_2+G^2_2 = 1$. When the measurement outcome is $b_1$, the 3-qubit state becomes
\begin{align}
\rho_{\hat{Y}_1}^{b_1}=&\frac{F_2}{2}\rho_{\hat{X}_2}^{a_2}+\frac{1+b_1G_2-F_2}{2}[U_{\hat{Y}_1}^{-1}\rho_{\hat{X}_2}^{a_2} (U_{\hat{Y}_1}^{-1})^\dagger] \nonumber\\ &+\frac{1-b_1G_2-F_2}{2}[U_{\hat{Y}_1}^{+1} \rho_{\hat{X}_2}^{a_2} (U_{\hat{Y}_1}^{+1})^\dagger]
\end{align}
where $ \hat{Y}_j=B_{j,m}= \vec{\xi}_{j,m} \cdot \vec{\sigma}$, $U_{\hat{Y}_j}^{b_j}=I\otimes \Pi_{\hat{Y}_j}^{b_j}\otimes I$. Similarly, if $\mathrm{Bob_2}$ performs a strong measurement $\hat{Y}_2$ with the outcome $b_2$, the 3-qubit state becomes $\rho_{\hat{Y}_2}^{b_2}=U_{\hat{Y}_2}^{b_2}\rho_{\hat{Y}_1}^{b_1} {U_{\hat{Y}_2}^{b_2}}^ \dagger$. Subsequently, $\mathrm{Charlie_1}$ performs a weak measurement $\hat{Z}_k$ on his received qubit with the quality factor $F_3$ and precision factor $G_3$ of the measurement, where the optimal weak measurement requires that $F^2_3+G^2_3 = 1$. When the measurement outcome is $c_1$, then the 3-qubit state changes to
\begin{align}
\rho_{\hat{Z}_1}^{c_1}=&\frac{F_3}{2}\rho_{\hat{Y}_2}^{b_2}+\frac{1+c_1G_3-F_3}{2}[U_{\hat{Z}_1}^{-1}\rho_{\hat{Y}_2}^{b_2}(U_{\hat{Z}_1}^{-1})^\dagger] \nonumber\\ &+\frac{1-c_1G_3-F_3}{2}[U_{\hat{Z}_1}^{+1}\rho_{\hat{Y}_2}^{b_2}(U_{\hat{Z}_1}^{+1})^\dagger]
\end{align}
where $ \hat{Z}_k=B_{k,n}= \vec{\xi}_{k,n} \cdot \vec{\sigma}$
and $U_{\hat{Z}_k}^{c_k}=I\otimes I\otimes \Pi_{\hat{Z}_k}^{c_k}$. Finally, when $\mathrm{Charlie_2}$ performs a strong measurement $\hat{Z}_2$ with the outcome $c_2$, the 3-qubit state will turn to $\rho_{\hat{Z}_2}^{c_2}=U_{\hat{Z}_2}^{c_2}\rho_{\hat{Z}_1}^{c_1} {U_{\hat{Z}_2}^{c_2}}^ \dagger$. So a cyclic measurement process of this scenario has been completely described. After repeating such process over and over again, we can obtain a complete joint probability distribution $P(a_{1},a_{2},b_{1},b_{2},c_{1},c_{2}\mid \hat{X}_{1},\hat{X}_{2},\hat{Y}_{1},\hat{Y}_{2},\hat{Z}_{1},\hat{Z}_{2})$. From the unnormalized postmeasurement state $\rho_{\hat{Z}_2}^{c_2}$, the joint conditional probability distribution is given as
$P(a_{1},a_{2},b_{1},b_{2},c_{1},c_{2}\mid \hat{X}_{1},\hat{X}_{2},\hat{Y}_{1},\hat{Y}_{2},\hat{Z}_{1},\hat{Z}_{2} )=\mathrm{Tr}[\rho_{\hat{Z}_2}^{c_2}]$.

Here we need to point out that only the two observers on the same side should measure sequentially, such as $\mathrm{Alice_1}$ should measure before $\mathrm{Alice_2}$, the same for $\mathrm{Bob_1}-\mathrm{Bob_2}$ and $\mathrm{Charlie_1}-\mathrm{Charlie_2}$. However, there is no assumption of measurement order between observers on different sides. In other words, the sequence of local measurements between the observers in three different sides does not change the final joint conditional probability distribution $P(a_{1},a_{2},b_{1},b_{2},c_{1},c_{2}\mid \hat{X}_{1},\hat{X}_{2},\hat{Y}_{1},\hat{Y}_{2},\hat{Z}_{1},\hat{Z}_{2} )$. For simplicity, we follow the sequence of $\mathrm{Alice_1}-\mathrm{Alice_2}-\mathrm{Bob_1}-\mathrm{Bob_2}-\mathrm{Charlie_1}-\mathrm{Charlie_2}$ to describe the measurement process.

\begin{figure}[htbp]
      \centering
      \includegraphics[width=0.4\textwidth]{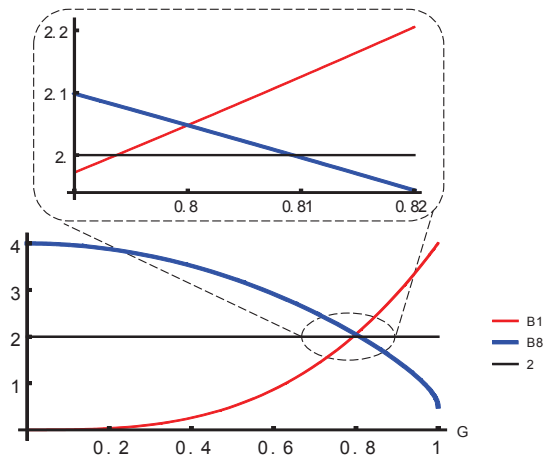}
      \caption{\small{Plot of MABK quantity $B_1$ for  $\mathrm{Alice_{1}-Bob_{1}-Charlie_{1}}$ and $B_8$ for $\mathrm{Alice_{2}-Bob_{2}-Charlie_{2}}$ when the state is GHZ state. The thin red line describes $B_1$ and the thick blue line describes $B_8$. When we choose $G_1$ = $G_2$ = $G_3$= $G$, they both exceed the bound of 2 in a narrow range. The picture above is the magnification of the violation part.}\label{2bob}}
    \end{figure}

According to the content discussed in this article, we are more concerned about the joint conditional probabilities for the measurement of any three different-side observers. Supposed that the observable choosing is unbiased for each observer, which requires that each measurement setting of every observer should be chosen with equal probability. The joint conditional probability $P(a_i,b_j,c_k\mid\hat{X}_{i},\hat{Y}_{j},\hat{Z}_{k})$ is obtained via
marginalizing the corresponding variables,
\begin{multline}
P(a_i,b_j,c_k\mid\hat{X}_{i},\hat{Y}_{j},\hat{Z}_{k})=\frac{1}{2^3}\sum_{\hat{X}_{i'},\hat{Y}_{j'},\hat{Z}_{k'}}\sum_{a_{i'},b_{j'},c_{k'}}\\
P(a_i,b_j,c_k,a_{i'},b_{j'},c_{k'}\mid\hat{X}_{i},\hat{Y}_{j},\hat{Z}_{k},\hat{X}_{i'},\hat{Y}_{j'},\hat{Z}_{k'}).
\end{multline}
Based on the joint conditional probability distribution, we can calculate the expected
value $E(\hat{X}_{i},\hat{Y}_{j},\hat{Z}_{k})$, which is given as
\begin{multline}\label{average}
E(\hat{X}_{i},\hat{Y}_{j},\hat{Z}_{k})=\sum_{a_i,b_j,c_k}a_ib_jc_k\ P(a_i,b_j,c_k\mid\hat{X}_{i},\hat{Y}_{j},\hat{Z}_{k}) \\
=\frac{1}{2^3}\sum_{a_i,b_j,c_k} a_ib_jc_k
\sum_{\hat{X}_{i'},\hat{Y}_{j'},\hat{Z}_{k'}} \sum_{a_{i'},b_{j'},c_{k'}} \\
P(a_i,b_j,c_k,a_{i'},b_{j'},c_{k'} \mid \hat{X}_{i},\hat{Y}_{j},\hat{Z}_{k},\hat{X}_{i'},\hat{Y}_{j'},\hat{Z}_{k'})
\end{multline}

The quantum non-locality can be witnessed via violations of corresponding inequalities. To explore the phenomenon of nonlocal sharing for a 3-qubit system via
multilateral sequential measurements, we will use the typical N-qubit Bell-type inequality, Mermin-Ardehali-Belinskii-Klyshko (MABK) inequality, which can be described as
\begin{eqnarray}\label{MABK}
&&|-E(A_{\mathrm{i,1}},B_{\mathrm{j,1}},C_{\mathrm{k,1}})+E(A_{\mathrm{i,2}},B_{\mathrm{j,1}},C_{\mathrm{k,2}})+\nonumber\\
&&E(A_{\mathrm{i,2}},B_{\mathrm{j,2}},C_{\mathrm{k,1}})+E(A_{\mathrm{i,1}},B_{\mathrm{j,2}},C_{\mathrm{k,2}})|\leq2.
\end{eqnarray}
Obviously, by choosing different observers in each side, we can discuss the violations of eight MABK inequalities. For clarity of discussions, we denote MABK quantity as $B_{\omega}$, which is the value on the left side of Eq.(\ref{MABK}) for the combination of different observers, where $B_1$ for $(i=j=k=1)$, $B_2$ for $(i=k=1,j=2)$, $B_3$ for $(i=j=1,k=2)$, $B_4$ for $(j=k=1,i=2)$,$B_5$ for$(j=k=2,i=1)$, $B_6$ for $(i=k=2,j=1)$, $B_7$ for $(i=j=2,k=1)$, $B_8$ for $(i=j=k=2)$. Each expected value in every MABK inequality can be obtained from Eq.(\ref{average}). Thus it is possible to check whether there exists multiple violation via calculation results, which will be analyzed in Sec. III.

\begin{figure}[htbp]
      \centering
      \includegraphics[width=0.45\textwidth]{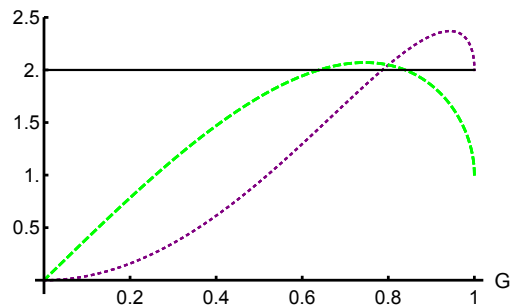}
      \caption{\small{Plot of MABK quantities $B_2$ to $B_7$ for $\mathrm{Alice-Bob-Charlie}$ when the state is GHZ state. When we choose $G_1$ = $G_2$ = $G_3$= $G$, $B_2$=$B_3$=$B_4$ (purple dotted line) and $B_5$=$B_6$=$B_7$ (green dashed line). They can exceed the classical bound simultaneously in a narrow range.}\label{2bob}}
 \end{figure}

\section{Nonlocal sharing in multi-sided Sequential Measurements Case}

As is known, non-locality sharing between $\mathrm{Alice_1-Bob_1}$ and $\mathrm{Alice_2-Bob_2}$ is impossible in a 2-qubit system \cite{Zhu}. We firstly explore whether nonlocal sharing between $\mathrm{Alice_{1}-Bob_{1}-Charlie_{1}}$ and $\mathrm{Alice_{2}-Bob_{2}-Charlie_{2}}$ exists in a 3-qubit system or not.

Without loss of generality, we assume that the observers in the three different sides share a three-qubit GHZ state, which is
\begin{eqnarray}\label{Bell state}
\mid\psi\rangle=\frac{1}{\sqrt{2}}(\mid 000\rangle+\mid 111\rangle).
\end{eqnarray}
Each observer has two measurement directions to choose, and the directions of the dichotomic measurements are denoted as $(\{\theta_{11},\phi_{11}\},\{\theta_{12},\phi_{12}\})$ for $\mathrm{Alice_1}$, $(\{\theta_{13},\phi_{13}\},\{\theta_{14},\phi_{14}\})$ for $\mathrm{Alice_2}$, $(\{\theta_{21},\phi_{21}\},\{\theta_{22},\phi_{22}\})$ for $\mathrm{Bob_1}$,
$(\{\theta_{23},\phi_{23}\},\{\theta_{24},\phi_{24}\})$ for $\mathrm{Bob_2}$, $(\{\theta_{31},\phi_{31}\},\{\theta_{32},\phi_{32}\})$ for $\mathrm{Charlie_1}$,
$(\{\theta_{33},\phi_{33}\},\{\theta_{34},\phi_{34}\})$ for $\mathrm{Charlie_2}$.

To explore nonlocal sharing between $\mathrm{Alice_{1}-Bob_{1}-Charlie_{1}}$ and $\mathrm{Alice_{2}-Bob_{2}-Charlie_{2}}$, it is necessary to determine whether the MABK quantities, $B_1$ and $B_8$, can surpass the classical bounds simultaneously or not. To simplify calculation, we require that the measurement directions of every observers will be always in the $X-Y$ plane, where $\theta_{11} =\theta_{12} =\theta_{13} =\theta_{14} =\theta_{21}=\theta_{22} =\theta_{23} =\theta_{24} =\theta_{31} =\theta_{32}=\theta_{33} = \theta_{34} = \frac{\pi}{2}$. Unfortunately, it is still too complex to obtain the optimal measurement settings which can show the maximal nonlocal sharing between $\mathrm{Alice_{1}-Bob_{1}-Charlie_{1}}$ and $\mathrm{Alice_{2}-Bob_{2}-Charlie_{2}}$. But we can easily show such nonlocal sharing by simple measurement settings even though they are suboptimal measurement settings. Interestingly, when we chose such simple measurement settings, $\phi_{11}=\phi_{21}=\phi_{31}=0$,
$\phi_{12}=\phi_{22}=\phi_{32}=-\phi_{14}=-\phi_{24}=\frac{\pi}{2}$, $\phi_{13}=-\phi_{23}=-\phi_{33}=\pi$, $\phi_{34}=\frac{3\pi}{2}$, the MABK quantities, $B_1$ and $B_8$, turn to
\begin{align}\label{MABK18}
B_1=&4G_1G_2G_3\\
B_8=&\frac{1}{2}(1+F_1)(1+F_2)(1+F_3).
\end{align}
For simplicity, when $G_1=G_2=G_3=G$, $B_1$ and $B_8$ changes to $B_1=4G^3$ and $B_8=\frac{1}{2}(1+\sqrt{1-G^2})^3$. The MABK quantities, $B_1$ and $B_8$, can exceed 2 simultaneously
in the narrow range of $G\in(\sqrt{2(2^{\frac{2}{3}}-2^{\frac{1}{3}})},2^{-\frac{1}{3}})$ (approximate $G\in(0.793,0.809)$). As illustrated in Fig. 2, when $G=0.8$, $B_1=B_8=2.048$, which is the maximal simultaneous violation for $B_1$ and $B_8$. Different from a 2-qubit case, it shows that non-locality sharing between $\mathrm{Alice_{1}-Bob_{1}-Charlie_{1}}$ and $\mathrm{Alice_{2}-Bob_{2}-Charlie_{2}}$ in 3-qubit system can be observed, where two MABK inequalities can be violated simultaneously.

Secondly, we can explore whether nonlocal sharing still exists or not for other different observers combinations. Certainly, it can be analyzed by discussing the simultaneous violation for these MABK quantities, $B_2$ to $B_7$.
When the same measurement directions mentioned above are used, the measurement directions of every observer will be always in the $X-Y$ plane, the MABK quantities, $B_2$ to $B_7$, can be written as
\begin{align}\label{MABK27}
B_2=&2(1+F_2)G_1G_3\nonumber\\
B_3=&2(1+F_3)G_1G_2\nonumber\\
B_4=&2(1+F_1)G_2G_3\nonumber\\
B_5=&(1+F_2)(1+F_3)G_1\nonumber\\
B_6=&(1+F_1)(1+F_3)G_2\nonumber\\
B_7=&(1+F_1) (1+F_2)G_3.
\end{align}
Similarly, when these MABK quantities in Eq.(\ref{MABK27}) can exceed 2 simultaneously, the nonlocal sharing phenomenon can be observed. For simplicity, when $G_1=G_2=G_3=G$, the MABK quantities $B_2$ to $B_4$ will change to the same value $2G^2(1+\sqrt{1-G^2})$, while the MABK quantities $B_5$ to $B_7$ will also change to another value $G(1+\sqrt{1-G^2})^2$. As illustrated in Fig. 3, it is easily to find the MABK quantities $B_2$ to $B_4$ will exceed 2 simultaneously in the range of $G\in(\sqrt{\frac{\sqrt{5}-1}{2}},1)$. When $G=\frac{2\sqrt{2}}{3}$, the MABK quantities $B_2$ to $B_4$ achieves the maximal value $2.37$ by choosing these measurement settings. The MABK quantities $B_5$ to $B_7$ will exceed 2 simultaneously in the range of $G\in(0.638,0.839)$, and the MABK quantities $B_5$ to $B_7$ achieves the maximal value $2.07$ by choosing these measurement settings when  $G=\frac{\sqrt{5}}{3}$. Hence, the MABK quantities $B_2$ to $B_7$ will exceed 2 simultaneously in the range of $G\in(\sqrt{\frac{\sqrt{5}-1}{2}},0.839)$. When $G=0.8$, $B_2=B_3=B_4=B_5=B_6=B_7=2.048$, which is the maximal simultaneous violation for $B_2$ to $B_7$.

Finally, when all the eight MABK quantities can exceed 2 simultaneously, it will exhibit a complete non-locality sharing in such a three qubit system via multilateral sequential measurements. Actually, the eight MABK inequalities, from $B_1$ to $B_8$, can be simultaneously violated. We can easily show such nonlocal sharing by simple measurement settings which are mentioned above, even though they are suboptimal measurement settings. As illustrated in Fig. 4, when $G_1=G_2=G_3=G$, we show all the eight MABK quantities will exceed 2 simultaneously in the range of $G\in(\sqrt{2(2^{\frac{2}{3}}-2^{\frac{1}{3}})},2^{-\frac{1}{3}})$. when $G=0.8$, $B_1=B_2=B_3=B_4=B_5=B_6=B_7=B_8=2.048$, which is the maximal simultaneous violation for all the eight MABK quantities. Compared with 2-qubit cases, the nonlocal sharing in a three qubit system shows more novel characteristics.

\begin{figure}[htbp]
      \centering
      \includegraphics[width=0.4\textwidth]{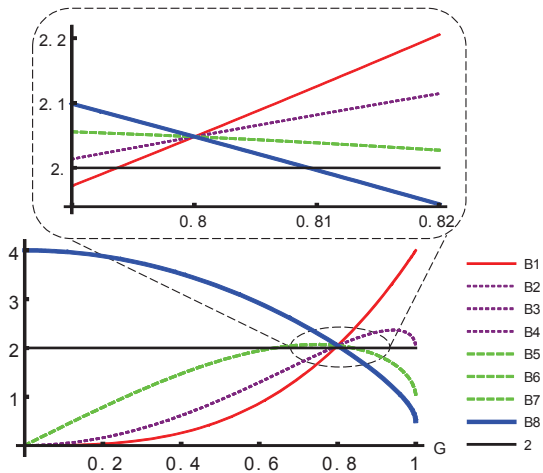}
      \caption{\small{Plot of MABK quantity $B_1$ to $B_8$ for $\mathrm{Alice-Bob-Charlie}$ when the state is GHZ state. When we choose $G_1$ = $G_2$ = $G_3$ = $G$, the expression become $B_2$=$B_3$=$B_4$ and $B_5$=$B_6$=$B_7$. The thin red solid line describes $B_1$, the purple dotted line describes $B_2$-$B_4$, the green dashed line describes $B_5$-$B_7$, the thick blue solid line describes $B_8$, and they both exceed the bound of 2 in a narrow range. The picture above is the magnification of the violation part.
}\label{2bob}}
    \end{figure}

\section{Nonlocal sharing perspective from joint conditional probability}

Although the investigation of nonlocal sharing scenario has made some progress, it is still in its infancy where lots of open questions are very confusing. As mentioned above, we explored the non-locality sharing for a three-qubit system via multilateral sequential measurements by investigating and analyzing the corresponding joint conditional probabilities. In this part of the discussion, we will try to give some beneficial viewpoints for nonlocal sharing from joint conditional probabilities.

When we try to investigate such nonlocal sharing scenario, we are easily troubled by such problems: why does such nonlocal sharing scenario is worth to explore? why did it not develop with the development of Bell's non-locality, but was ignored for a long time?

Not only Bell's non-locality, but also the whole EPR quantum correlation problem can be summarized as ``A thought caused by a real conditional probability ''. In fact, no matter which kind of quantum correlation criterion, inequality type, entropy criterion, geometric criterion, etc., all can be expressed in the form of the joint conditional probabilities.

If we understand them from the perspective of joint conditional probabilities, we note that almost all the criteria and experimental verification involved in quantum correlation can be described as follows: for a quantum state system with $N$ particles, $N$ observers will share and measure them. Each observer may have multiple measurement settings, but every observer will only measure the particle in its hand once in each round of measurement. After repeating such process over and over again, the joint conditional probability distribution based on $N$ measurements will be obtained. Taking a 2-qubit system as example, all the criteria can start from a joint conditional probability $P(a,b\mid\hat{A},\hat{B},\rho)$. Then, the combination of these conditional probabilities or the physical quantities which (average value, etc.) are based on the conditional probabilities are used to exhibit the conflict between quantum prediction and classical model. One of the key features in this scenario is that each shared particle of the quantum state will be measured once and discarded. However, this feature should not be an inevitable condition. On the contrary, the conditional probability can be derived from a higher dimensional conditional probability based on more measurements. Consider the simplest example, for a 2-qubit system, we assumed that each particle will be measured twice in each round. For instance, Alice will carry out sequential measurements of $\hat{A}_{1}$,$\hat{A}_{2}$ and Bob will carry out sequential measurements of $\hat{B}_{1}$,$\hat{B}_{2}$. Then a joint conditional probability $P({a}_{1},{a}_{2},{b}_{1},{b}_{2}\mid\hat{A}_{1},\hat{A}_{2},\hat{B}_{1},\hat{B}_{2},\rho)$ can be obtained. The nonclassical correlation in a 2-qubit system will be expressed more comprehensive by using this joint conditional probability, as the previous joint conditional probability can be always obtained via marginal constraints, $P({a}_{1},{b}_{1}\mid\hat{A}_{1},\hat{B}_{1},\rho)=\sum_{{a}_{2},{b}_{2}}P({a}_{1},{a}_{2},{b}_{1},{b}_{2}\mid\hat{A}_{1},\hat{A}_{2},\hat{B}_{1},\hat{B}_{2},\rho)$.

However, previous studies ignored the research path based on the above joint conditional probability. The crucial factor is about the measurements. In previous scenario, the observer's measurements are usually strong measurements. Taking a 2-qubit system scenario as an example, Alice and Bob preform strong measurement in their first measurement, which means $\hat{A}_{1}$,$\hat{B}_{1}$ are strong measurements. Even if the subsequential second measurements are carried out, the joint conditional probability can be always written as $P({a}_{1},{a}_{2},{b}_{1},{b}_{2}\mid\hat{A}_{1},{A}_{2},{B}_{1},{B}_{2},\rho)=P({a}_{1},{b}_{1}\mid{A}_{1},{B}_{1},\rho)P({a}_{2}\mid\hat{A}_{2},\Pi_{\hat{A}_1}^{a_1})P({b}_{2}\mid\hat{B}_{2},\Pi_{\hat{B}_1}^{b_1})$,  $\Pi_{\hat{A}_1}^{a_1}$ and $\Pi_{\hat{B}_1}^{b_1}$ are the eigenstates of $\hat{A}_{1}$ and $\hat{B}_{1}$. In this scenario, the later measurement results will not carries any correlation information of the initial state, so it is meaningless to discuss sequential measurements. However, if the decomposition of the conditional probability can not take the equal sign, the problem will become not trivial. From the physical point of view, in order to satisfy this requirement, it is necessary to protect the correlation information of the initial state as much as possible in the former measurements. Fortunately, weak measurement as an important nonperturbative measurements satisfies such a condition. Nonlocal sharing emerges in such scenario. Hence the observation of nonlocal sharing usually associate with weak measurement or POVM measurement.


\section{Conclusion}

The phenomenon of non-locality sharing for a 3-qubit system via multilateral sequential measurements has been completely discussed. In order to compare with a 2-qubit case, we firstly explored non-local sharing in a 3-qubit system between $\mathrm{Alice_1-Bob_1-Charlie_1}$ and $\mathrm{Alice_2-Bob_2-Charlie_2}$. Interestingly, the corresponding MABK inequalities, $B_1$ and $B_8$, can exceed 2 simultaneously in the narrow range of $G\in(\sqrt{2(2^{\frac{2}{3}}-2^{\frac{1}{3}})},2^{-\frac{1}{3}})$. Hence, nonlocal sharing in a 3-qubit system between $\mathrm{Alice_1-Bob_1-Charlie_1}$ and $\mathrm{Alice_2-Bob_2-Charlie_2}$ can be observed, while it is impossible for a 2-qubit case. Secondly, we also investigated nonlocal sharing for the other different observers combinations. It is shown that the MABK quantities $B_2$ to $B_7$ will exceed 2 simultaneously in the range of $G\in(\sqrt{\frac{\sqrt{5}-1}{2}},1)$. Finally, all the eight possible MABK inequalities in this scenario were fully explored. Actually, the eight MABK inequalities, from $B_1$ to $B_8$, can be violated simultaneously. We can easily show such nonlocal sharing by simple measurement settings which are mentioned above, even though they are suboptimal measurement settings. When $G_1=G_2=G_3=G$, we show all the eight MABK quantities will exceed 2 simultaneously in the range of $G\in(\sqrt{2(2^{\frac{2}{3}}-2^{\frac{1}{3}})},2^{-\frac{1}{3}})$. when $G=0.8$, $B_1=B_2=B_3=B_4=B_5=B_6=B_7=B_8=2.048$, which is the maximal simultaneous violation for all the eight MABK quantities. Compared with a 2-qubit case, the nonlocal sharing in a 3-qubit system shows more novel characteristics. Besides, a deep understanding of the phenomenon of nonlocal sharing from the view of joint conditional probabilities were exhibited, which can give some beneficial viewpoints for deeply understanding general nonlocal sharing phenomenon.

\section{Acknowledgment}
C.R. was supported by the National Natural Science Foundation of China (Grant No. 12075245), National Key Research and Development Program (Grant No. 2017YFA0305200), and Xiaoxiang Scholars Programme of Hunan Normal University. X.Z. was supported by the National Key Research and Development Program (2017YFA0305200 and 2016YFA0301700), the Key Research and Development Program of Guangdong Province of China (2018B030329001 and 2018B030325001), the National Natural Science Foundation of China (Grant No.61974168), the Natural Science Foundation of Guangdong Province of China (2016A030312012), and the National Young 1000 Talents Plan.


\end{document}